\documentclass[a4paper,10pt,twoside]{cpc-hepnp}

\usepackage{multicol}
\usepackage{graphicx}
\usepackage{booktabs}
\usepackage{amssymb,bm,mathrsfs,bbm,amscd}
\usepackage[tbtags]{amsmath}
\usepackage{lastpage}
\usepackage{color}

\begin{document}

\fancyhead[c]{\small Chinese Physics C~~~Vol. 39, No. 10 (2015) 104001}
\fancyfoot[C]{\small 104001-\thepage}

\footnotetext[0]{Received 19 March 2015}

\title{Direct mass measurements of neutron-rich $^{86}$Kr projectile fragments and the persistence of neutron magic number $N$ = 32 in Sc isotopes  \thanks{This work is supported in part by the 973 Program of China (No. 2013CB834401), the NSFC (Grant Nos. U1232208, U1432125, 11205205, 11035007) and the Helmholtz-CAS Joint Research Group (Group No. HCJRG-108).}}

\author{%
      XU Xing$^{1,2,3}$
\quad WANG Meng$^{1;1)}$\email{wangm@impcas.ac.cn}%
\quad ZHANG Yu-hu$^{1;2)}$\email{yhzhang@impcas.ac.cn}%
\quad XU Hu-shan$^{1}$
\quad SHUAI Peng$^{4}$
\quad TU Xiao-lin$^{1,5}$\\
\quad Yuri A. Litvinov$^{1,5}$
\quad ZHOU Xiao-hong$^{1}$
\quad SUN Bao-hua$^{6}$
\quad YUAN You-jin$^{1}$
\quad XIA Jia-wen$^{1}$
\quad YANG Jian-cheng$^{1}$\\
\quad Klaus Blaum$^{7}$
\quad CHEN Rui-jiu$^{1}$
\quad CHEN Xiang-cheng$^{1,2}$
\quad FU Chao-yi$^{1,2}$
\quad GE Zhuang$^{1,2}$
\quad HU Zheng-guo$^{1}$\\
\quad HUANG Wen-jia$^{1,2}$
\quad LIU Da-wei$^{1,2}$
\quad LAM Yi-hua$^{1}$
\quad MA Xin-wen$^{1}$
\quad MAO Rui-shi$^{1}$
\quad T. Uesaka$^{9}$\\
\quad XIAO Guo-qing$^{1}$
\quad XING Yuan-ming$^{1,2}$
\quad T. Yamaguchi$^{8}$
\quad Y. Yamaguchi$^{9}$
\quad ZENG Qi$^{4}$
\quad YAN Xin-liang$^{1}$\\
\quad ZHAO Hong-wei$^{1}$
\quad ZHAO Tie-cheng$^{1}$
\quad ZHANG Wei$^{1,2}$
\quad ZHAN Wen-long$^{1}$
}
\maketitle

\address{%
$^1$ Key Laboratory of High Precision Nuclear Spectroscopy and Center for Nuclear Matter Science, Institute of Modern Physics, Chinese Academy of Sciences, Lanzhou 730000, China\\
$^2$ University of Chinese Academy of Sciences, Beijing, 100049, China\\
$^3$ CSNSM-IN2P3-CNRS, Universit\'{e} de Paris Sud, F-91405 Orsay, France\\
$^4$ Research Center for Hadron Physics, National Laboratory of Heavy Ion Accelerator Facility in Lanzhou and University of Science and Technology of China, Hefei 230026, China \\
$^5$ GSI Helmholtzzentrum f\"{u}r Schwerionenforschung, Planckstra{\ss}e 1, 64291 Darmstadt, Germany\\
$^6$ School of Physics and Nuclear Energy Engineering, Beihang University, Beijing 100191, China\\
$^7$ Max-Planck-Institut f\"{u}r Kernphysik, Saupfercheckweg 1, 69117 Heidelberg, Germany\\
$^8$ Department of Physics, Saitama University, Saitama 338-8570, Japan\\
$^9$ RIKEN Nishina Center, RIKEN, Saitama 351-0198, Japan

}

\begin{abstract}
In this paper, we present direct mass measurements of neutron-rich $^{86}$Kr projectile fragments conducted at the HIRFL-CSR facility in Lanzhou by employing the Isochronous Mass Spectrometry (IMS) method. The new mass excesses of $^{52-54}$Sc nuclides are determined to be -40492(82), -38928(114), -34654(540) keV, which show a significant increase of binding energy compared to the reported ones in the Atomic Mass Evaluation 2012 (AME12). In particular, $^{53}$Sc and $^{54}$Sc are more bound by 0.8 MeV and 1.0 MeV, respectively. The behavior of the two neutron separation energy with neutron numbers indicates a strong sub-shell closure at neutron number  $N$ = 32 in Sc isotopes.
\end{abstract}



\footnotetext[0]{\hspace*{-3mm}\raisebox{0.3ex}{$\scriptstyle\copyright$}2013
Chinese Physical Society and the Institute of High Energy Physics
of the Chinese Academy of Sciences and the Institute
of Modern Physics of the Chinese Academy of Sciences and IOP Publishing Ltd}%

\begin{multicols}{2}

\section{Introduction}

Atomic nuclei are many-body quantum systems composed of two distinct types of fermions: protons and neutrons.
Some special nuclei with certain configurations of protons and neutrons are found to be more bound and more stable than others~\cite{Elsasser}. These magic nuclei are
cornerstones of the nuclear shell model which was introduced by Mayer and Jense more than 60 years ago~\cite{Mayer}. Since then, how magic numbers evolve with
extreme proton-to-neutron ratios toward the drip lines has become one of the research frontiers of nuclear physics. The nuclear mass reflects the total complex effect of strong, weak and electromagnetic interactions among nucleons \cite{Klaus,Lunney}, and thus is a useful tool for this research.

Recently, studies of exotic nuclei, especially neutron-rich nuclei, far away from the $\beta$-stability line have resulted in some fruit.
The $^{24}$O nucleus has been verified to be doubly magic by three different experiments~\cite{24O1,24O2,24O3}, confirming the appearance of the new magic number $N$ = 16. On the
other hand, several mass measurements of neutron-rich Si, P, S and Cl isotopes~\cite{28N1,28N2,28N3,28N4} indicate the reduction of the traditional $N$ = 28 neutron shell gap. This appearance or collapse of the magic number is often believed to be caused by the change of the single-particle level order~\cite{order}.

In $fp$-shell nuclei, the magicity of $N$ = 32 has been substantially found in several experiments with Ca~\cite{ca}, Ti~\cite{ti}, and Cr~\cite{cr} isotopes by investigating the
systematic behavior of the $E$($2^{+}_1$) energies with the neutron number in the even-even nuclei of the corresponding isotopic chain. The new magic number $N$ = 32 in Ca isotopes is also confirmed
by recent precision mass measurements of  $^{51,52}$Ca at the TITAN Penning trap facility~\cite{TITANPRL} and $^{53,54}$Ca with the multi reflection time-of-flight mass spectrometer of ISOTRAP at CERN~\cite{ISOLDE}.

In this paper, we report on the results from isochronous mass measurement of $^{86}$Kr projectile fragments in the vicinity of $N$ = 32 at the Institute of Modern Physics in Lanzhou,
China. Most of our results are in good agreement with the literature ones reported in AME12~\cite{AME1,AME2}, except for four nuclides: $^{52-54}$Sc and $^{56}$Ti. Masses of all nuclides of interest are re-evaluated. With the new data, an
$N$ = 32 neutron shell gap energy in Sc isotopes is obtained, which is conspicuously stronger than the former one determined from previous data.

\section{Experiment and Data Analysis}

This experiment was performed at the HIRFL-CSR accelerator facility~\cite{Zhan10}. The primary beam of $^{86}$Kr$^{28+}$ ions was accumulated in the main Cooler Storage Ring (CSRm)
and then accelerated to an energy of 460.65 Mev/u. The $^{86}$Kr$^{28+}$ ions were fast extracted and then focused on a
$\sim$15 mm thick beryllium target which was placed at the entrance of the RIBLL2 (an in-flight fragment separator). The secondary ions were produced via projectile fragmentation of $^{86}$Kr$^{28+}$. Exotic fragments were separated by the RIBLL2
and then injected to the experimental Cooler Storage Ring (CSRe). Both RIBLL2 and CSRe were set to a fixed magnetic rigidity of $B\rho$ = 7.6755 Tm for optimum transmission of $^{61}$Cr$^{24+}$.
Other fragments within a $B\rho$-acceptance of about $\pm$0.2\% of the RIBLL2-CSRe system were also transmitted and stored.
As a result, about 5 ions were simultaneously stored in CSRe in each injection.

For various ions stored in the CSRe, their revolution times $T$ are a function of their mass-to-charge ratios $m/q$ and their velocities $v$ in the first order approximation as follows~\cite{RadonPRL,HauNIMA_IMS,Franzke}:
\begin{equation}
\frac{\Delta{T}}{T} \approx \frac{1}{\gamma_t^2} \frac{\Delta(m/q)}{m/q}-(1-\frac{\gamma^2}{\gamma_t^2})\frac{\Delta{v}}{v},\label{eq1}
\end{equation}
where $\gamma_t$ is the so-called transition point of the ring \cite{Franzke,BOSCH} and $\gamma$ is the the relativistic Lorentz factor. In the IMS experiment, the ring is tuned to a special
ion-optical mode~\cite{Xia_NIMA2002_488} allowing the faster ions of a certain ion species always to circulate in longer orbits while the slower ones are in correspondingly shorter orbits. The energy of the primary beam was chosen such that the isochronous condition $\gamma \approx \gamma _t$ is fulfilled. Hence, the velocity spread of injected ions
is compensated by their orbit lengths and thus the revolution times directly reflect $m/q$ ratios of the stored ions~\cite{TUXLPRL_11}.

At each revolution, the stored ions passed through a dedicated timing detector \cite{MEIBO} equipped with
a 19 $\mu$g/cm$^2$ thin carbon foil of 40 mm in diameter installed inside CSRe. When passing through, the ions lose energy so that secondary electrons are released from this foil.
These electrons were collected by a set of micro-channel plates, thus forming a time stamp for each revolution. The revolution times
of all stored ions used in further analysis were extracted from these time stamps. More details can be found in Refs~\cite{TUNIMA,XUHS}.

This experiment lasted for about five days. The instabilities of the CSRe magnetic fields caused drifts of the entire revolution times and thus limited the achievable mass resolving power for
all ions. We applied a new data analysis method, as described in Ref~\cite{SHUAINIMA}, to minimize the effect of such instabilities.
By accumulating the results of each injection and carefully correcting the effect of instability of the magnetic field, a common revolution time spectrum was obtained, from which nuclides were identified applying the method described in Refs \cite{TUNIMA,HauNIMA_IMS}.
Fig. \ref{sepectra} shows a revolution time spectrum of most nuclides within a range of ${603}$ ns ${\leq}$ t ${\leq}$ ${622}$ ns.
\begin{center}
\includegraphics[width=8cm]{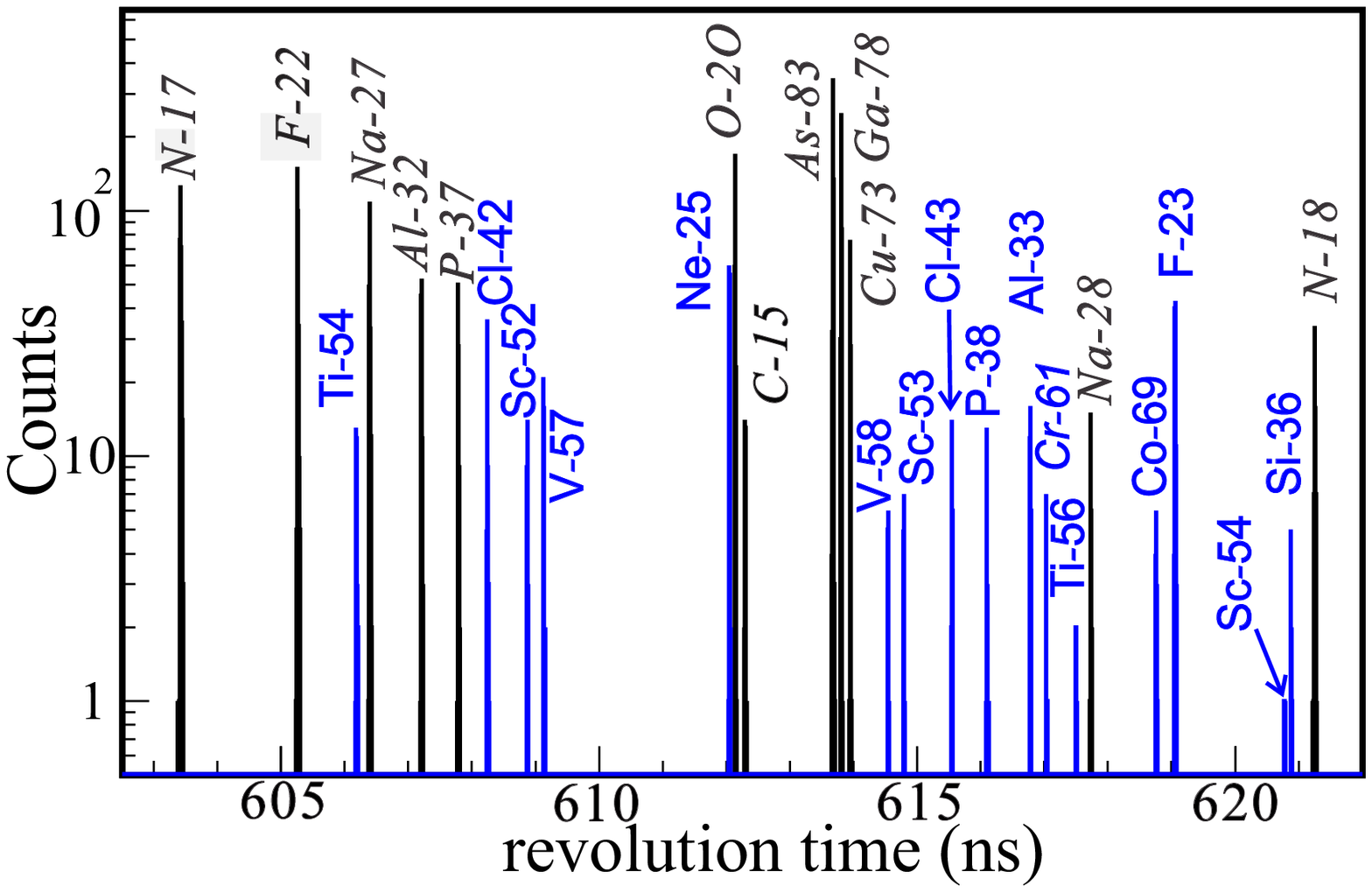}
\figcaption{Corrected revolution time spectrum of $^{86}$Kr fragments stored in CSRe. Nuclei with masses determined in this
experiment and those used as references are indicated with bold
and italic letters, respectively.
 } \label{sepectra}
\end{center}

To calibrate the mass values of stored ions from that spectrum, a third polynomial function was utilized to describe the relationship between $m/q$ ratios and revolution times $T$. In this
calibration, twelve nuclides with accurately known masses~\cite{AME2} were selected as reference to deduce the freely fitted parameters of Eq.(\ref{eq2}), and thus masses of interested nuclides can be obtained by interpolating the fitting function to the corresponding revolution time $T$.
\begin{equation}
\frac{m}{q}(T)=a_0+a_1\cdot{T}+a_2\cdot{T^2}+a_3\cdot{T^3}.
\label{eq2}
\end{equation}

In order to evaluate the reliability of our results, we re-determined the masses of each of the twelve reference
nuclides by calibrating the spectrum with the other eleven nuclides. The comparison between the re-determined mass values and those in the literature
are presented in Fig.~\ref{calibrant}. The normalized $\chi _n$ value was defined as:
\begin{equation}
\chi_n=\sqrt{\frac{1}{n} \sum_{i=1}^{n}{\frac{(ME_{CSRe,i}-ME_{AME,i})^2}{\sigma_{CSRe,i}^2+\sigma_{AME,i}^2}}}
\label{eq3}
\end{equation}
with $n$ = 12 in our case. The obtained value of $\chi _n$ = 1.45 is slightly out of the expected range of $\chi _n$ = $1$  $\pm$ $0.24$ at the 1$\sigma$ confidence level, which means
that an additional systematic error $\sigma_{sys}$ = $35$ keV must be required.

\begin{center}
\includegraphics[width=8cm]{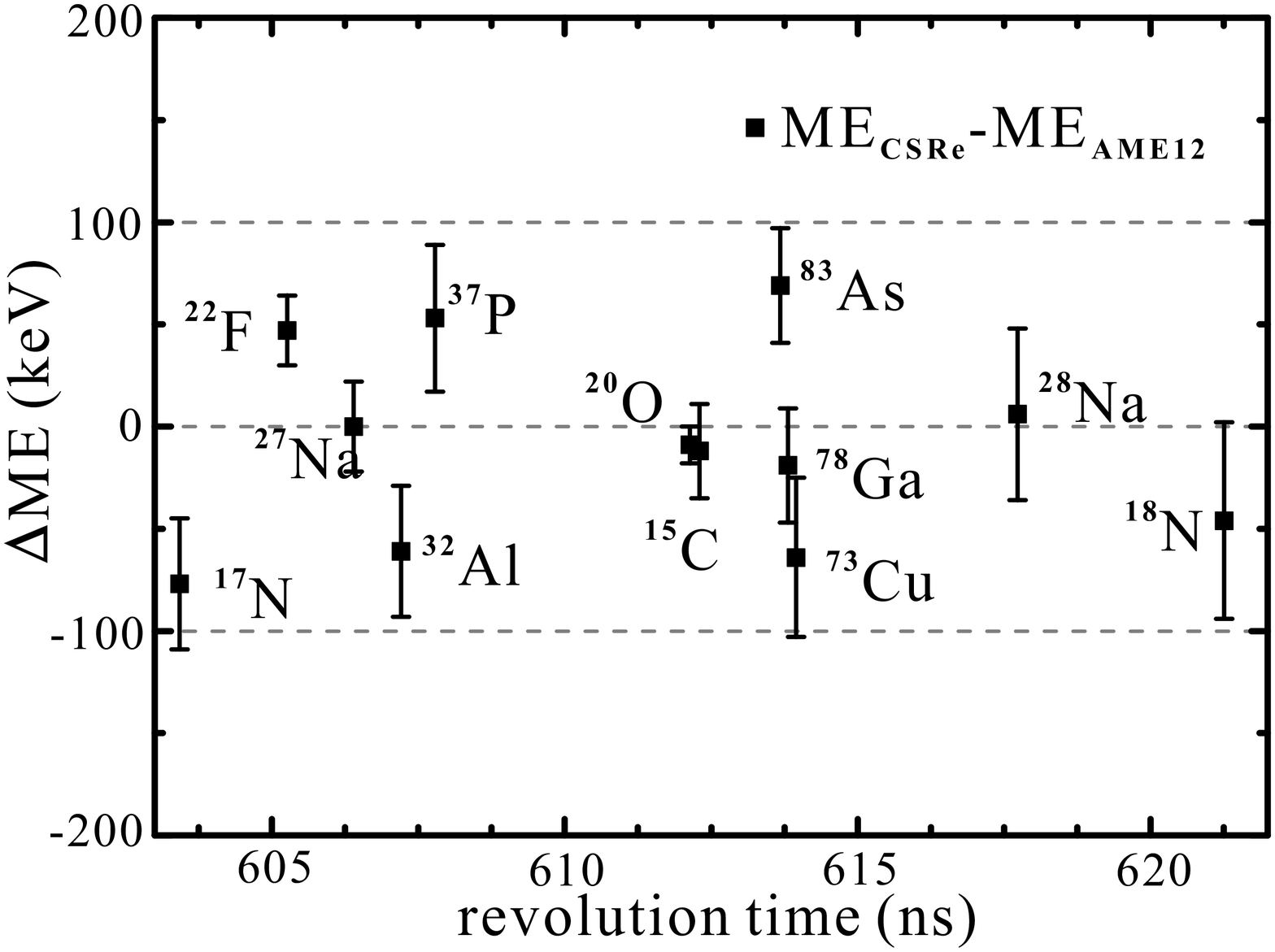}
\figcaption{ Comparison of mass excess (ME) values re-determined in this work with those in AME12. Note that the error bars for each nuclide in this figure contain only the statistical error.
 } \label{calibrant}
\end{center}

\section{Results and discussion}

Our new mass excess (ME) values of nuclides of interest as well as corresponding ones reported in AME12~\cite{AME2} are presented in Table~\ref{tab1}. Most of our results are in good agreement with those in the literature at the 1$\sigma$ confidence level except for four nuclides $^{52-54}$Sc and $^{56}$Ti. Our ME values of these four nuclides are smaller by $1.8\sigma$, $2.8\sigma$, $1.6\sigma$ and $1.9\sigma$ than the corresponding ones in AME12.
For nuclides with consistent mass values, our results are included in a new atomic mass evaluation.  The re-evaluated mass values are the weighted averages of our ME values and corresponding values from the literature and are presented in Table~\ref{tab1}.

For the four nuclides with observed deviations, we investigated further and found that there were six corresponding direct mass measurements. ME values of the four nuclides from these experiments are presented in Table~\ref{tab2}.

For $^{53}$Sc and $^{54}$Sc, results from CSRe, GSI \cite{MATOS} and TOFI2 \cite{TOFI2} are in excellent agreement with each other. The results from TOFI1 \cite{TOFI1} show that the two nuclides are more than 1 MeV unbound compared to the three former experiments. This is similar to the situation for $^{52}$Ca, for which an increase of 1.74 MeV in binding energy was discovered by the precise Penning-trap measurement established at TITAN~\cite{TITANPRL}.
 As a consequence, we think that the results from TOFI1 might need to be revised. The results from TOFI3 \cite{TOFI3}, although with large uncertainties, also disagree for $^{53}$Sc. However, AME12 only adopted the recent results of the two nuclides from MSU~\cite{MSUPRL}. The MSU data show that the two nuclides are more unbound by about 0.8 MeV and 1.0 MeV  compared to the three  experiments which support each other. Moreover, for $^{53}$Sc, the MSU data agree with the result from TOFI1 but not with TOFI3, while the conclusion for $^{54}$Sc was the reverse. Hence, the choice for the two nuclides in AME12 seems to be questionable. In our re-evaluation, we employed results from CSRe, GSI and TOFI2 and the re-evaluated values are displayed in Table~\ref{tab2}.

For $^{52}$Sc, the results from CSRe and three TOFI experiments agree well with each other. However, we still only adopt results from CSRe and TOFI2 to be consistent with the choice for $^{53}$Sc and $^{54}$Sc.

For $^{56}$Ti, the situation is more complicated, as even the results from CSRe, GSI and TOFI2 disagree slightly with each other. We have to ignore the dispute, and give a suggested mass excess value adopting the results from all three chosen experiments as mentioned above. More precise mass experiments for this nuclide are necessary to shed light on this conflict.

\begin{center}
\tabcaption{ \label{tab1}  Mass excess values in keV of nuclides of interest from this work and the AME12 literature values. The given mass excess uncertainties from our work contain the statistical and systematic errors. The re-evaluated values are displayed in the last column.}
\footnotesize
\begin{tabular}{p{0.7cm} p{1.5cm} p{1.5cm} p{1.35cm} p{1.6cm}}
\toprule Atom~ & ~~ME$_{CSRe}$  & ME$_{AME12}$& $~~\Delta$ME & re-evaluated \\
\hline
$^{23}$F~~~   &  ~~$3263(45)$ & ~~$3310(50)$  & $-57(68)$     & ~~$3284(34)$  \\
$^{25}$Ne~~   & $-2018(38)$   & $-2060(48)$   & ~~$42(62)$    & $-2034(30)$ \\
$^{33}$Al~~   & $-8578(58)$   & $-8470(80)$   & $-108(99)$    & $-8540(47)$  \\
$^{36}$Si~~   & $-12560(152)$ & $-12390(70)$  & $-170(167)$   & $-12420(64)$ \\
$^{38}$P~~~   & $-14613(90)$  & $-14670(90)$  & ~~$57(127)$   & $-14642(64)$ \\
$^{42}$Cl~~   & $-24832(60)$  & $-24910(144)$ & ~~$78(156)$   & $-24844(55)$\\
$^{43}$Cl~~   & $-24164(65)$  & $-24320(100)$ & ~~$156(120)$  & $-24210(55)$ \\
\textcolor[rgb]{1.00,0.00,0.00}
{$^{52}$Sc}~~   & $-40462(90)$   & $-40170(140)$ & \textcolor[rgb]{1.00,0.00,0.00}{$-292(167)$} &  \\
\textcolor[rgb]{1.00,0.00,0.00}
{$^{53}$Sc}~~   & $-38940(115)$  & $-38110(270)$ & \textcolor[rgb]{1.00,0.00,0.00}{$-830(294)$} & \\
\textcolor[rgb]{1.00,0.00,0.00}
{$^{54}$Sc}~~   & $-34653(545)$  & $-33600(360)$ & \textcolor[rgb]{1.00,0.00,0.00}{$-1053(654)$} & \\
$^{54}$Ti~~     & $-45641(110)$  & $-45600(120)$ & $-41(163)$  & $-45622(81)$ \\
\textcolor[rgb]{1.00,0.00,0.00}
{$^{56}$Ti}~~   & $-39775(270)$  & $-39210(140)$ & \textcolor[rgb]{1.00,0.00,0.00}{$-565(305)$} & \\
$^{57}$V~~     & $-44435(85)$  & $-44230(230)$ & $-205(245)$  & $-44410(80)$ \\
$^{58}$V~~     & $-40480(125)$  & $-40320(130)$ & $-160(180)$  & $-40400(90)$ \\
$^{61}$Cr~    & $-42508(164)$  & $-42460(130)$ & $-48(210)$  & $-42480(102)$ \\
$^{69}$Co~    & $-50410(208)$  & $-50170(190)$ & $-240(282)$  & $-50280(140)$ \\
\bottomrule
\end{tabular}
\vspace{0mm}
\end{center}
\vspace{0mm}
\end{multicols}

\begin{center}
\tabcaption{ \label{tab2} Mass excess (ME) values in keV of $^{52-54}$Sc and $^{56}$Ti from present work, an isochronous mass measurement at GSI~\cite{MATOS}, three independent mass measurements at TOFI ~\cite{TOFI1,TOFI2,TOFI3}, and a TOF-B$\rho$ measurement at MSU~\cite{MSUPRL}. One can see AME12~\cite{AME1} to have a quick review of these data.}
\footnotesize
\begin{tabular*}{170mm}{@{\extracolsep{\fill}}cccccccc}
\toprule Atom &   ME$_{\rm CSRe}$ &ME$_{\rm GSI}$      &ME$_{\rm TOFI1}$~\cite{TOFI1}  &ME$_{\rm TOFI2}$~\cite{TOFI2}   &ME$_{\rm TOFI3}$~\cite{TOFI3}   &ME$_{\rm MSU}$   &re-evaluated  \\
\hline
$^{52}$Sc     & $-40462(90)$  &                & $-40520(220)$ & $-40380(230)$ & $-40150(225)$  &               & $-40450(85)$ \\
$^{53}$Sc     & $-38940(115)$ & $-38840(110)$  & $-38600(250)$ & $-38970(260)$ & $-38290(370)$  & $-38110(270)$ & $-38895(76)$ \\
$^{54}$Sc     & $-34653(545)$ & $-34520(210)$  & $-33500(500)$ & $-34520(465)$ & $-34430(370)$  & $-33540(360)$ & $-34535(180)$  \\
$^{56}$Ti     & $-39775(270)$ & $-39420(120)$  & $-38470(350)$ & $-39130(280)$ & $-38900(250)$  &               & $-39435(105)$  \\
\bottomrule
\end{tabular*}%
\end{center}

 \begin{multicols}{2}
 The mass excesses or binding energies reflect the total effect of interactions within nuclei and provide essential information about the ordering of single-particle energies in exotic regions of the
 nuclear chart. As can be seen in Fig.~\ref{s2n}, our results show the decrease in mass excess leads to a distinct variation of the two neutron separation energy $S_{2n}$ in the vicinity of $N$ = 32 and $Z$ = 20. $S_{2n}$ is defined as $S_{2n}=ME(N,Z-2)+2ME(n)-ME(N,Z)$, where $N$ and $Z$ are neutron numbers and proton numbers, respectively. To better understand the behavior of the $S_{2n}$ with increasing neutron number in this region, new $^{53,54}$Ca data from ISOTRAP ~\cite{ISOLDE} are also included.
 \begin{center}
\includegraphics[width=8cm]{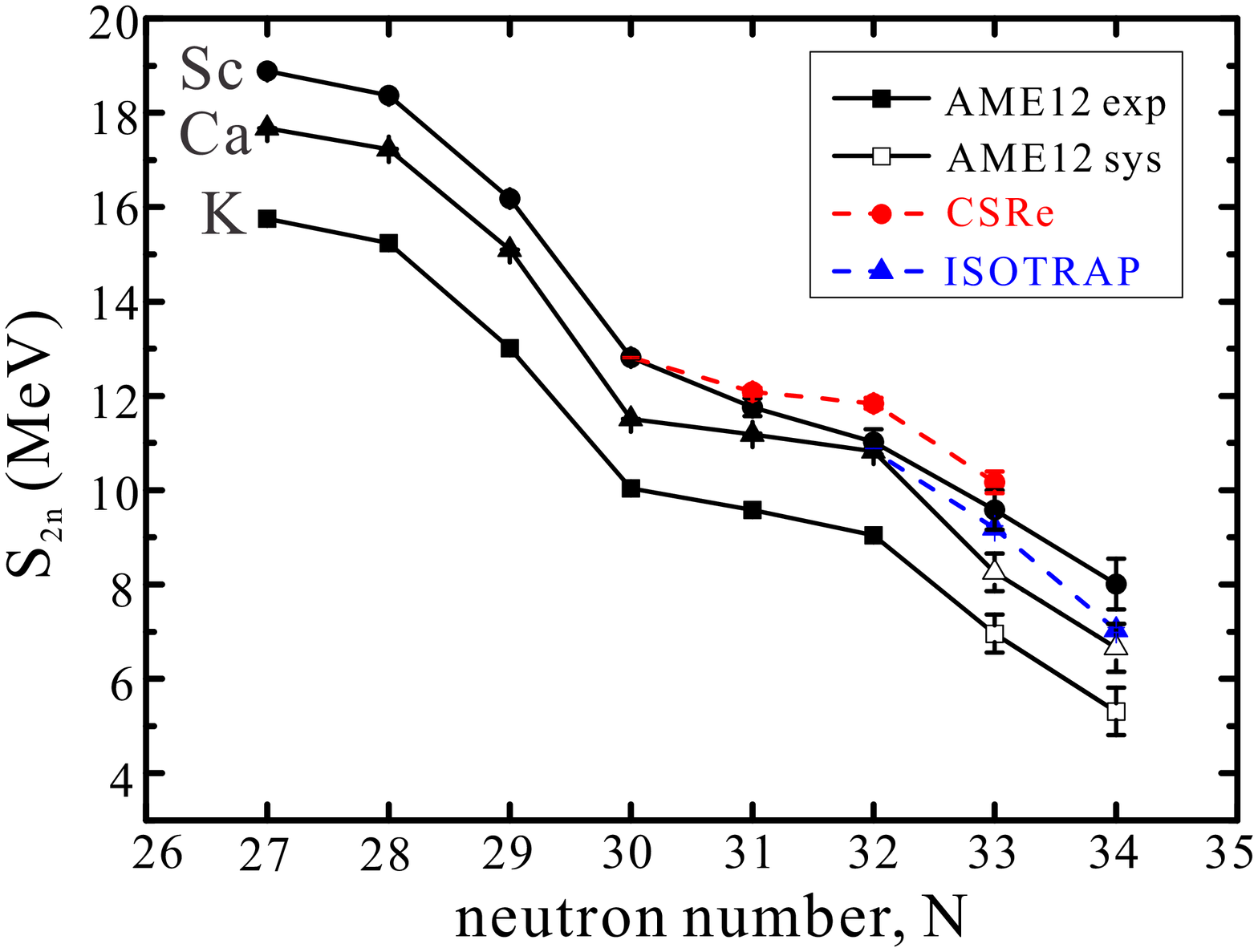}
\figcaption{Two-neutron separation energy $S_{2n}$ as a function of neutron number $N$ for the K (squares),
Ca (triangles), and Sc (circles) isotopic chains. Solid symbols connected by solid lines represent values based completely on experimental data from AME12 while hollow ones are estimated in AME12.
Red solid circles connected by red dashed lines are derived from mass values from our work, and blue solid triangles are derived from $^{53,54}$Ca mass values from Ref~\cite{ISOLDE}.
 } \label{s2n}
\end{center}

The consequent behavior of $S_{2n}$ with increasing neutron number in the Sc isotopic chain is dramatically flatter from $N$ = 30 to $N$ = 32. This is in line with the behavior in the K and Ca isotopic chains, as discussed in Refs~\cite{TITANPRL,ISOLDE} and explained theoretically by the effect of three nucleon (3N) forces.

 From $N$= 32 to $N$ = 33, the drop of $S_{2n}$ occurs in both Ca and Sc isotopic chains and is as steep as the drop from $N$ = 28 to $N$ = 29. The latter manifests the effect of the traditional neutron magic number 28. Hence, the former is substantial evidence of the persistence of $N$ = 32 magicity in the Sc isotopes. Although this magicity has been well established by many experiments as introduced in the introduction, it is confirmed in an odd-Z isotopic chain for the first time.
\begin{center}
\includegraphics[width=8cm]{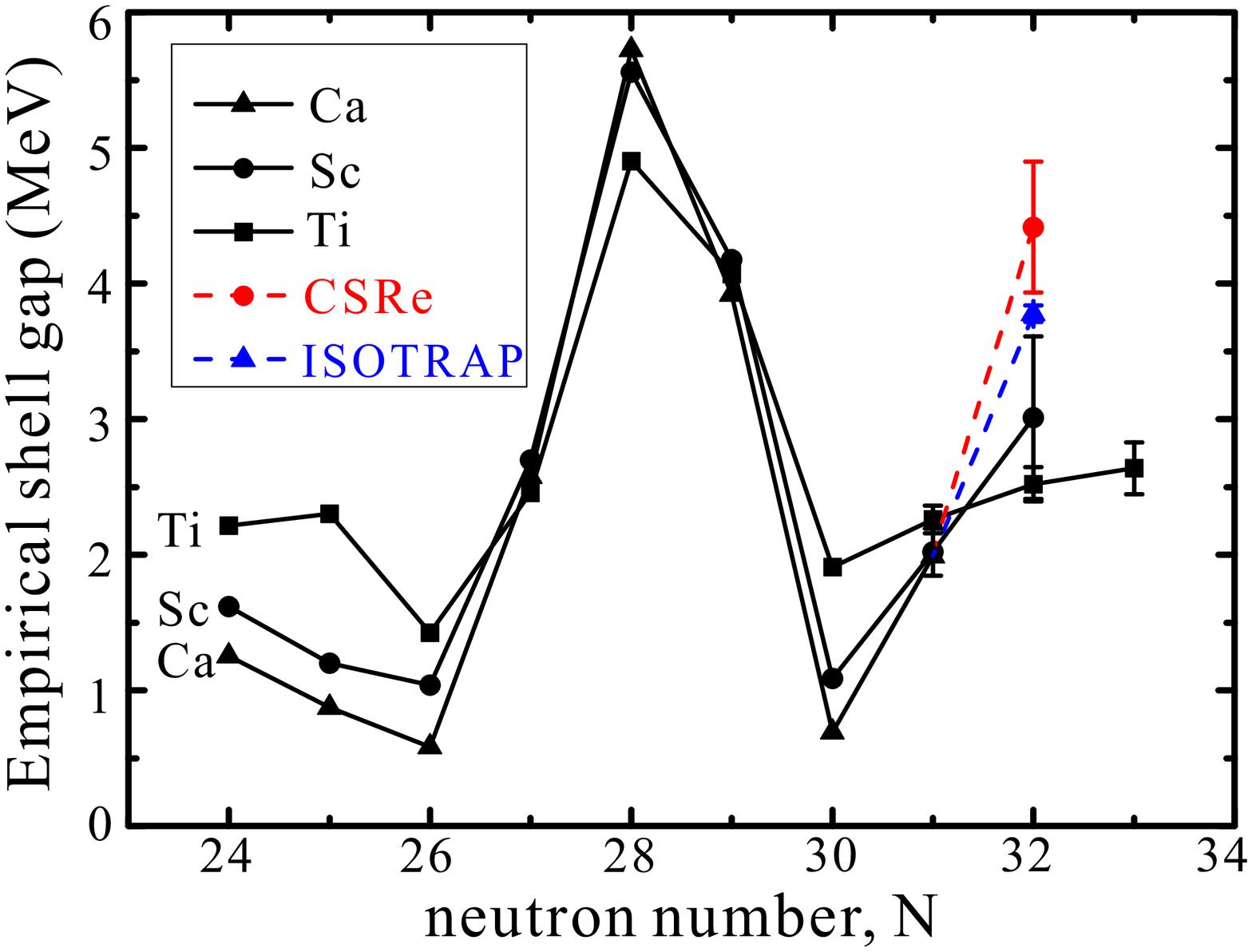}
\figcaption{Empirical shell gap energy as a function of neutron number $N$ for the Ca (triangles),
Sc (circles), and Ti (squares) isotopic chains. Solid symbols connected by solid lines represent values based completely on experimental data from AME12.
The red solid circle is derived from $^{53}$Sc mass values from our work and $^{51,55}$Sc from AME12, while the blue solid triangle is derived from $^{50,52}$Ca mass values from AME12 and $^{54}$Ca from Ref~\cite{ISOLDE}.
 } \label{gap}
\end{center}


 The strength of this sub-shell closure can be evaluated via the neutron shell gap energy, defined as the difference of two neutron separation energy $\Delta(N,Z)=S_{2n}(N,Z)-S_{2n}(N+2,Z)$. Fig.~\ref{gap} illustrates the systematic behavior of empirical neutron shell gap energy with neutron number in Ca, Sc, and Ti isotopic chain. The distinct peak of neutron shell gap energy at $N$ = 28 demonstrates the traditional neutron magic number 28 again. One can see that the shell gap for $^{52}$Ca is almost 4 MeV, which unambiguously establishes a prominent shell closure at $N$ = 32 in Ca isotopes~\cite{ISOLDE}. For $^{53}$Sc, the shell gap value obtained from our data is more than 4 MeV and is about 2 MeV stronger than the former one obtained from AME12, although these two values both have large uncertainties.

\section{Summary}

The results of an isochronous mass measurement of $^{86}$Kr projectile fragments are presented in this paper. Our results show that $^{53}$Sc and $^{54}$Sc are stonger bound by  0.8 MeV and 1.0 MeV than the literature values reported in AME12, respectively. However, our results agree perfectly with two other experiments, which are not adopted in AME12. It seems that the data selections in AME12 for these two nuclides are not very reasonable. Mass values of the nuclides of interest
are re-evaluated including our new data. The large increase in binding energies of $^{53}$Sc and $^{54}$Sc suggests the persistence of the new neutron magic number $N$=32 in Sc isotopes, which has already been confirmed in the Ca isotopic chain by several experiments. The physics behind this phenomena is still under exploration.

\acknowledgments{We thank Professor Georges Audi for very helpful discussion about Atomic Mass Evaluation and data analysis.}

\end{multicols}

\vspace{15mm}

\vspace{-1mm}
\centerline{\rule{80mm}{0.1pt}}
\vspace{2mm}

\begin{multicols}{2}

\end{multicols}

\clearpage


\begin{thebibliography}{90}

\vspace{3mm}
\bibitem{Elsasser} ELSASSER W, J. de Phys. et Rad., 1934, {\bf 5}: 625.
\bibitem{Mayer} GOEPPERT MAYER M, Phys. Rev. 1949, {\bf 75}: 1969-1970.
\bibitem{Klaus} BLAUM K,  Phys. Rep., 2006, {\bf 425}: 1-78.
\bibitem{Lunney} LUNNEY D  et al. Rev.~Mod.~Phys.~{\bf 75} (2003) 1021.
\bibitem{24O1} OZAWA A et al. Phys. Rev. Lett., 2000, {\bf 84}: 5493-5495.
\bibitem{24O2} HOFFMAN C R et al. Phys. Rev. Lett., 2008, {\bf 100}: 152502.
\bibitem{24O3} KANUNGO R et al. Phys. Rev. Lett., 2009, {\bf 102}: 152501.
\bibitem{28N1} SAVAJOLS H et al. Eur. Phys. J. A, 2005, {\bf 25}: 23.
\bibitem{28N2} SARAZIN F et al. Phys. Rev. Lett., 2000, {\bf 84}: 5062.
\bibitem{28N3} JURADO B et al. Phys. Lett. B, 2007, {\bf 649}: 43.
\bibitem{28N4} RINGLE R et al. Phys. Rev. C, 2009, {\bf 80}: 064321.
\bibitem{order} OTSUKA T et al. Phys. Rev. Lett., 2001, {\bf 87}: 082502
\bibitem{ca} GADE A et al. Phys. Rev. C, 2006, {\bf 74}: 021302(R).
\bibitem{ti} JANSSENS R V F et al. Phys. Lett. B, 2002, {\bf 546}: 55.
\bibitem{cr} PRISCIANDARO J I et al. Phys. Lett. B, 2001, {\bf 510}: 17.
\bibitem{TITANPRL} GALLANT A T et al. Phys. Rev. Lett., 2012, {\bf 109}: 032206.
\bibitem{ISOLDE} WIENHOLTZ F et al. Nature, 2013, {\bf 498}: 346-349.
\bibitem{AME1} AUDI G, WANG M et al. Chin. Phys. C, 2012, {\bf 36}: 1287-1602.
\bibitem{AME2} WANG M, AUDI G et al. Chin. Phys. C, 2012, {\bf 36}: 1603-2014.
\bibitem{MSUPRL} ESTRADE A et al. Phys. Rev. Lett., 2011, {\bf 107}: 172503.
\bibitem{Zhan10} ZHAN W L et al. Nucl. Phys. A, 2010, {\bf 834}: 694c.
\bibitem{RadonPRL} RADON T, KERSCHER T, SCHLITT B et al. Phys. Rev. Lett., 1997, {\bf 78}:4701.
\bibitem{HauNIMA_IMS} HAUSMANN M et al. Nucl. Instr. and Meth. in Phys. Res. A, 2002, {\bf 488}: 11-25.
\bibitem{Franzke} FRANZKE B, GEISSEL H, M\"{U}NZENBERG. Mass. Spectrom. Rev., 2008, {\bf 27}:428-469.
\bibitem{BOSCH} BOSCH F, LITVINOV Y A, ST\"{O}HLKER T Prog. in Par. and Nucl. Phys, 2013, {\bf 73} 84-140.
\bibitem{Xia_NIMA2002_488} XIA J W, ZHAN W L, WEI B W et al. Nucl. Instr. and Meth. in Phys. Res. A, 2002, {\bf 488}: 11-25.
\bibitem{TUXLPRL_11}  TU X L, XU H S, WANG M et al. Phys. Rev. Lett., 2011, {\bf 106}: 112501.
\bibitem{MEIBO} MEI B, TU X L, WANG M et al. Nucl. Instr. and Meth. in Phys. Res. A, 2010, {\bf 624}: 109-113.
\bibitem{TUNIMA} TU X L, WANG M, LITVINOV Y A et al. Nucl. Instr. and Meth. in Phys. Res. A, 2011, {\bf 654}: 213-218.
\bibitem{XUHS} XU H S, ZHANG Y H, LITVINOV Y A Int. J. Mass Spectrom., 2013, {\bf 349-350}: 162-171.
\bibitem{SHUAINIMA} SHUAI P et al. arXiv:1407.3459
\bibitem{MATOS} MATOS M, Ph.D. thesis, Justus-Liebig-Universitat Giessen, 2004.
\bibitem{TOFI1} TU X L et al. Z. Phys. A Atomic Nuclei, 1990, {\bf 337}: 361-366.
\bibitem{TOFI2} SEIFERT H L et al. Z. Phys. A Atomic Nuclei, 1994, {\bf 349}: 25-31.
\bibitem{TOFI3} BAI Y et al. AIP Conf. Proc, 1998, {\bf 455}: 90.
\end{thebibliography}
\end{document}